\documentstyle[preprint,aps,epsf]{revtex}
\newcommand {\bea}{\begin{eqnarray}}
\newcommand {\eea}{\end{eqnarray}}
\newcommand {\be}{\begin{equation}}
\newcommand {\ee}{\end{equation}}
\newcommand {\qslash}{q\!\!\!/}
\newcommand {\muslash}{\mu\!\!\!/}

\begin{document}

%\draft
\preprint{IASSNS-HEP 99/58}

\title{Superconductivity from perturbative one-gluon
exchange in high density quark matter}

\author{Thomas Sch\"afer\footnote{Research supported in part by NSF
PHY-9513835.  e-mail: schaefer@sns.ias.edu. Present address: TRIUMF,
4004 Wesbrook Mall, Vancouver, BC, Canada V6T 2A3.}
and Frank Wilczek\footnote{Research supported in part by DOE grant
DE-FG02-90ER40542. e-mail: wilczek@sns.ias.edu }}

\address{School of Natural Sciences\\
Institute for Advanced Study\\
Princeton, NJ 08540}

\maketitle

\begin{abstract}
  We study color superconductivity in QCD at asymptotically
large chemical potential. In this limit, pairing is dominated
by perturbative one-gluon exchange. We derive the Eliashberg
equation for the pairing gap and solve this equation numerically.
Taking into account both magnetic and electric gluon exchanges,
we find $\Delta\sim g^{-5}\exp(-c/g)$ with $c=3\pi^2/\sqrt{2}$,
verifying a recent result by Son. For chemical potentials that
are of physical interest, $\mu< 1$ GeV, the calculation ceases to 
be reliable quantitatively,
but our results suggest that the gap can be as 
large as 100 MeV.
\end{abstract}

\newpage

  1. The behavior of matter at very high baryon density but small
temperature is of interest in connection with the physics of
neutron stars and heavy ion collisions in the baryon rich regime. 
Moreover, it has been realized that matter at very high 
density exhibits many non-perturbative phenomena, such as a mass
gap and chiral symmetry breaking, in a regime where the coupling 
is weak and systematic calculations are possible. 

  At very high density the natural starting point is a Fermi
sphere of quarks. The corresponding low energy excitations 
are quasiparticles and holes in the vicinity of the Fermi
surface. Since the Fermi momentum is large, asymptotic freedom
implies that the interaction between quasiparticles is weak. 
However, as we know from the theory of superconductivity, 
the Fermi surface is unstable in the presence of even an
arbitrarily weak attractive interaction. In QCD, the attraction
is provided by one-gluon exchange between quarks in a 
color anti-symmetric $\bar 3$ state. QCD at high density 
is therefore expected to be a color superconductor
\cite{Frau_78,Bar_77}.

  A particularly interesting case is QCD with three
flavors. In this case the most favorable type of pairing
involves the coupling of color and flavor degrees of freedom, 
color-flavor-locking \cite{ARW_98b}. This implies, among 
other things, that all gluons acquire a mass and that 
chiral symmetry is broken. We have argued that in the 
color-flavor-locked phase not only universal features,
in particular the symmetry breaking pattern, but also many
non-universal properties, such as the spectrum of low-lying
states, exactly match the expectations for hadronic matter
at low baryon density \cite{SW_98b}. This means that 
nuclear matter at low density might be continuously 
connected to quark matter at high density, without 
any phase transition. 

  It also means that at very high density many interesting
properties of hadronic matter, such as the magnitude of 
the chiral condensate, can be calculated in weak coupling
perturbation theory. The first step in such a program is
the calculation of the superconducting gap. This calculation
was first attempted by Bailin and Love \cite{BL_84}. They
used a schematic IR cutoff in the gluon propagator. In their
treatment, the gap is of the form $\Delta\sim\mu\exp(-c/g^2)$, 
where $c$ depends logarithmically on the IR cutoff. Recently,
the problem was revisited by Son \cite{Son_98}, who argued
that the gap is dominated by magnetic gluon exchanges, and
that the infrared behavior is regulated by dynamic screening.
He obtained $\Delta\sim\mu g^{-5}\exp(-3\pi^2/(\sqrt{2}g))$.

  Our purpose in the present work is to rederive and strengthen 
this result, and to determine the overall numerical coefficient.

  2. In order to derive a gap equation, we follow the standard
Nambu-Gorkov formalism and introduce a two component field 
$\Psi=(\psi,\bar\psi^T)$. The inverse quark propagator takes 
the form
\be
\label{sinv}
S^{-1}(q) = \left(\begin{array}{cc}
 \qslash+\muslash-m &  \bar\Delta \\
 \Delta  & (\qslash-\muslash+m)^T 
\end{array}\right),
\ee
where $\bar\Delta=\gamma_0\Delta^\dagger\gamma_0$. 
The gap is a matrix in color, isospin, and Dirac space. 
In the following we consider the case of two massless
flavors. We will assume that the gap is anti-symmetric 
in both flavor and color, and has total angular momentum
zero. In the case of short range interactions, this 
assumption can be justified from our study of the 
renormalization group equations for a general four-fermion
interaction \cite{SW_98,EHS_98}. The one-gluon exchange
interaction is long range, and other forms of pairing 
might take place. In particular, since the interaction 
is dominated by almost collinear scattering, one might 
expect higher partial waves to play a role \cite{Son_98}. 
In the following, we will concentrate on the total
angular momentum zero gap.

 We also assume that the gap has positive parity. 
One-gluon exchange does not distinguish between pairs 
of positive or negative parity \cite{PR_98}. This degeneracy 
is lifted by instantons, which favor the positive parity
channel \cite{ARW_98,RSSV_98,SW_98}. At large chemical 
potential instanton effects are exponentially suppressed. 
In the following, we will therefore assume that the only
instanton effect is to determine the parity of the gap. 

 In addition to that, we neglect quark mass effects and
chiral symmetry breaking $LR$ condensates. As shown in 
\cite{SW_98} there is no BCS instability in the case 
of pairing between left and right handed quarks. The
formation of $LR$ condensates is therefore suppressed
by $m/\mu$. The form of the gap matrix is then 
\cite{BL_84,PR_99}
\be
\label{gap_ans}
 \Delta^{ab}_{ij}(q) = (\lambda_2)^{ab}(\tau_2)_{ij}
  C\gamma_5 \left( \Delta_1(q_0)\frac{1}{2}(1+\vec\alpha\cdot\hat q )
             +\Delta_2(q_0)\frac{1}{2}(1-\vec\alpha\cdot\hat q ) \right).
\ee
In the weak coupling limit, we can replace $\vec\alpha\cdot\hat q$
by the unit matrix using the equations of motion. In this limit, only 
$\Delta_1$ survives. We will see this explicitly from the solution 
of the gap equation presented below. We have neglected the dependence
of the gap on the magnitude of the 
momentum, but kept the dependence on frequency. 
The dependence on momentum can be dropped because, in the 
weak coupling limit, all momenta are close to the Fermi 
surface. For short range interactions, the dependence on
frequency can also be neglected. This is not the case here.
Because long range interactions are important, retardation
effects cannot be neglected. 

 The self energy in the Nambu-Gorkov formalism obeys the 
Dyson-Schwinger equation \cite{BL_84}
\be
\label{ds}
 \Sigma(k) = -ig^2 \int \frac{d^4q}{(2\pi)^4}
  \Gamma_\mu^a S(q)\Gamma_\nu^b D^{ab}_{\mu\nu}(q-k).
\ee
Here, $\Sigma(k)=-(S^{-1}(k)-S_0^{-1}(k))$ is the proper 
self energy, $\Gamma^a_\mu$ is the quark-gluon vertex
and $D^{ab}_{\mu\nu}(q-k)$ is the gluon propagator. To 
leading order in the perturbative expansion, we can use
the free vertex
\be
\label{vert_0}
\Gamma^a_\mu = \left(\begin{array}{cc}
 \gamma_\mu\lambda^a/2 & 0 \\
 0 & -(\gamma_\mu\lambda^a/2)^T \end{array}\right).
\ee
We will study the importance of vertex corrections
below. To leading order, we can also neglect the 
diagonal part of the proper self energy, that is the
fermion wave function renormalization \cite{Son_98}. 
In this case, (\ref{ds}) reduces to an equation 
for the gap matrix, 
\be 
\label{gap2}
 \Delta(k) = ig^2 \int \frac{d^4q}{(2\pi)^4}
 \left(\gamma_\mu\frac{\lambda^a}{2}\right)^T
 S_{21}(q) \left(\gamma_\nu\frac{\lambda^a}{2}\right)
 D_{\mu\nu}(q-k).
\ee  
Here, $S_{21}(q)$ is the 21-component of the fermion 
propagator in the Gorkov representation. $S_{21}(q)$
is determined from the inverse of (\ref{sinv}). We 
have
\be
\label{S21}
 S_{21}(q) = -\frac{1}{(\qslash-\muslash)^T}\Delta
   \frac{1}{(\qslash+\muslash) +\bar\Delta
     [(\qslash-\muslash)^T]^{-1}\Delta}.
\ee
Inserting the ansatz (\ref{gap_ans}) for the gap gives
\be 
\label{S21_2}
 S_{21}(q) = -\frac{1}{2}(\lambda_2\tau_2C\gamma_5) \left(
 \frac{\Delta_1(1-\vec\alpha\cdot\vec q)}{q_0^2-(|\vec{q}|-\mu)^2-\Delta_1^2} 
 + \frac{\Delta_2(1+\vec\alpha\cdot\vec q)}{q_0^2-(|\vec{q}|+\mu)^2-
\Delta_2^2}
 \right).
\ee 
Both the RHS and the LHS of the gap equation are proportional
to $\tau_2$, so the flavor structure simply drops out. The 
color coefficient is given by
\be
\label{color}
c = \frac{1}{4}(\lambda_a)^T\lambda_2\lambda_a
  = -\frac{N_c+1}{2N_c}\lambda_2 = -\frac{2}{3}\lambda_2 
 \hspace{0.5cm}(N_c=3),
\ee
where we have used the Fierz identity $(\lambda^a)_{ij}
(\lambda^a)_{kl}=-2/N_c\delta_{ij}\delta_{kl}+2\delta_{il}
\delta_{jk}$ and the factor 1/4 comes from the color generators 
$t^a=\lambda^a/2$. Projecting (\ref{gap2}) on $\Delta_{1,2}$ gives 
two coupled gap equations
\bea
\label{gap3}
\Delta_{1,2}(k_0) &=& \frac{2ig^2}{3} \int \frac{d^4q}{(2\pi)^4}
  \left\{ \frac{1}{8}{\rm tr}
  \left(\gamma_\mu(1-\vec\alpha\cdot\hat q)
        \gamma_\nu(1\pm\vec\alpha\cdot\hat k)\right)
  \frac{\Delta_1(q_0)}{q_0^2-(|\vec{q}|-\mu)^2-\Delta_1(q_0)^2}
  \right. \nonumber \\
  & & \hspace{1cm}\left. + \frac{1}{8}{\rm tr}
 \left(\gamma_\mu(1+\vec\alpha\cdot\hat q)
        \gamma_\nu(1\pm\vec\alpha\cdot\hat k)\right)
  \frac{\Delta_2(q_0)}{q_0^2-(|\vec{q}|+\mu)^2-\Delta_2(q_0)^2}
  \right\} D_{\mu\nu}(k-q),
\eea
where the two signs of $\vec\alpha\cdot\hat k$ on the 
RHS correspond to $\Delta_1$ and $\Delta_2$ on the LHS. 

 We now must specify the gluon propagator. The gluon propagator 
in a general covariant gauge is given by
\be
\label{D_dec}
 D_{\mu\nu}(q) = \frac{P_{\mu\nu}^T}{q^2-G} 
 + \frac{P_{\mu\nu}^L}{q^2-F} - \xi\frac{q_\mu q_\nu}{q^4}
\ee
where $D$ and $F$ are functions of $q_0$ and $|\vec{q}|$ and
the projectors $P_{\mu\nu}^{T,L}$ are defined by 
\bea
\label{proj}
 P_{ij}^T &=& \delta_{ij}-\hat{q}_i\hat{q}_j , \hspace{1cm}
 P_{00}^T = P_{0i}^T = 0,    \\
 P_{\mu\nu}^L &=& -g_{\mu\nu}+\frac{q_\mu q_\nu}{q^2}
   -P_{\mu\nu}^T .
\eea
It contains the gauge parameter $\xi$, which must not appear
in physical results.
In the weak coupling limit, $q_0$ is small as compared to 
$|\vec{q}|$. In this case we can expand the projectors 
$P_{\mu\nu}^L\simeq \delta_{\mu 0}\delta_{\nu 0}$ and 
$q_i q_j/q^2\simeq \hat{q}_i\hat{q}_j$. The gap equation 
now becomes
\bea
\label{gap_4}
\Delta_{1}(k_0) &=& -\frac{2ig^2}{3} \int \frac{d^4q}{(2\pi)^4}
  \left\{
  \frac{\Delta_1(q_0)}{q_0^2-(|\vec{q}|-\mu)^2-\Delta_1(q_0)^2}
  \left(
  \frac{\frac{3}{2}-\frac{1}{2}\hat{k}\cdot\hat{q}}{(k-q)^2-G}
 +\frac{\frac{1}{2}+\frac{1}{2}\hat{k}\cdot\hat{q}}{(k-q)^2-F}
  \right)
  \right. \nonumber \\
  & & \hspace{1.cm}\left. + 
  \frac{\Delta_2(q_0)}{q_0^2-(|\vec{q}|+\mu)^2-\Delta_2(q_0)^2}
 \left(
  \frac{\frac{1}{2}+\frac{1}{2}\hat{k}\cdot\hat{q}}{(k-q)^2-G}
 +\frac{\frac{1}{2}-\frac{1}{2}\hat{k}\cdot\hat{q}}{(k-q)^2-F}
 +\frac{\xi}{(q-k)^2}
  \right)
  \right\}.
\eea
There is a similar equation for $\Delta_2$ in which the two
terms in the round brackets are interchanged. Only the first
term term in (\ref{gap_4}) has a singularity on the Fermi
surface. In the weak coupling limit, we can therefore drop 
the second term, and we are left with an equation for 
$\Delta(p_0)\equiv \Delta_1(p_0)$. This equation is 
independent of the gauge parameter $\xi$. The second
gap parameter $\Delta_2$ is not suppressed in magnitude.
However, $\Delta_2$ does not lead to a gap on the Fermi 
surface, and its value is gauge dependent. 

  We should note that the fact that the gap is gauge 
independent in the present 
weak-coupling approximation is a consequence 
of the fact that the gap is determined by the scattering
of quarks that are almost on shell. For on-shell quarks,
the fact that the gauge dependent part of the propagator
does not contribute follows directly from the equations
of motion for the quark fields. 

  For large chemical potential the integral over $q$ is 
dominated by momenta in the vicinity of the Fermi surface,
$|\vec{q}|\simeq\mu$ and $q_0\ll\mu$. We can expand all
momenta as $\vec{q}=\vec{q}_F+\vec{l}$, where $\vec{q}_F$
is on the Fermi surface, and $\vec{l}$ is orthogonal to it.
Asymptotically, $|\vec{l}|\ll|\vec{q_F}|$ and the integration 
measure becomes $dq_0\,\mu^2 dl\,d\cos\theta\,d\phi$. We
also have $|\vec{q}-\vec{k}|\simeq\sqrt{2}\mu (1-\cos\theta)$.
The integral over $\phi$ is performed trivially. We analytically
continue to imaginary $q_0$, and perform the integral over 
$\vec{l}$ by picking up the pole in the diquark propagator. 
We find
\bea
\label{gap_5}
\Delta(p_0) &=& \frac{g^2}{12\pi^2} \int dq_0\int d\cos\theta\,
 \left(\frac{\frac{3}{2}-\frac{1}{2}\cos\theta}
            {1-\cos\theta+(G+(p_0-q_0)^2)/(2\mu^2)}\right. \\
 & & \hspace{3cm}\left.    +\frac{\frac{1}{2}+\frac{1}{2}\cos\theta}
            {1-\cos\theta+(F+(p_0-q_0)^2)/(2\mu^2)} \right)
 \frac{\Delta(q_0)}{\sqrt{q_0^2+\Delta(q_0)^2}}. \nonumber
\eea
The integral over $\cos\theta$ is dominated by small $\theta$, 
corresponding to almost collinear scattering. It is therefore 
important to take medium modifications of the gluon propagator
at small momenta into account. For $q_0\ll\vec{q}\to 0$ and to 
leading order in perturbation theory we have
\be
 F = 2m^2, \hspace{1cm}
 G = \frac{\pi}{2}m^2\frac{q_0}{|\vec{q}|},
\ee
with $m^2=N_fg^2\mu^2/(4\pi^2)$. In the longitudinal part,
$m_D^2=2m^2$ is the familiar Debye screening mass. In the 
transverse part, there is no screening of static modes, 
but nonstatic modes are modes are dynamically screened
due to Landau damping. In our case, typical frequencies
are on the order of the gap, $q_0\simeq \Delta$. This means
that the electric part of the interaction is screened at
$q_E\simeq m_D^{1/2}$ whereas the magnetic interaction
is screened at $q_M\simeq (\pi/4\cdot m_D^2\Delta)^{1/3}$.

 Asymptotically, $q_M\ll q_E$, and magnetic gluon exchange
dominates over electric gluon exchange. We therefore begin
by analyzing the gap equation taking into account the 
magnetic part of the interaction only. We will also 
approximate $\cos\theta\simeq 1$ in the denominator
and drop $(q_0-p_0)^2$ in the denominator. All of these
terms will be reinstated later. The integration over 
$\cos\theta$ is now straightforward. We have
\be
\label{eliash}
\Delta(p_0) = \frac{g^2}{18\pi^2} \int dq_0
 \log\left(1 + \frac{64\pi\mu}{N_fg^2|p_0-q_0|}\right) 
 \frac{\Delta(q_0)}{\sqrt{q_0^2+\Delta(q_0)^2}}.
\ee
If we are only interested in the leading exponential 
behavior of the gap we can drop the numerical factors 
and the powers of $g$ in the logarithm. We then arrive
at
\be
\label{eliash1}
\Delta(p_0) = \frac{g^2}{18\pi^2} \int dq_0
 \log\left(\frac{\mu}{|p_0-q_0|}\right) 
 \frac{\Delta(q_0)}{\sqrt{q_0^2+\Delta(q_0)^2}},
\ee
which is the equation discussed in the appendix of Son's
paper \cite{Son_98}. This equation was derived from the
on-shell quark-quark scattering amplitude. What we have 
shown here is that one can indeed derive this equation 
from the Dyson-Schwinger equation in the weak coupling
limit, and that the result is independent of the gauge 
parameter. Son also derives an approximate solution to 
this integral equation, 
\be
\label{sol_son}
\Delta_{\rm app.}(p_0) \equiv \Delta_0 \sin\left(\frac{g}{3\sqrt{2}\pi}
 \log\left(\frac{\mu}{p_0}\right)\right),\hspace{0.5cm}
 p_0>\Delta_0,
\ee
with $\Delta_0=\mu\exp(-3\pi^2/(\sqrt{2}g))$. The approximations
involved are expected to reproduce the correct coefficient in
the exponent, but do not fix the prefactor. 

 3. We have therefore solved the Eliashberg equation (\ref{eliash1})
numerically for different chemical potentials. We have used 
the one-loop running coupling constant evaluated at the Fermi
momentum $p_F=\mu$. This is an average over the momenta of the 
exchanged gluons, which are in the range $[q_M,2\mu]$.  Without 
a higher order calculation one cannot fix the scale in the running
coupling. We will see that the preexponential factor in the final
result behaves as $g^{-5}$. This factor is almost optimal to
give a remarkably weak scale dependence. 

 The result for the function $\Delta(p_0)$ for $\mu=400$ 
MeV and $\mu=10^{10}$ MeV is shown in Fig. \ref{fig_gap}. The 
solid line is the numerical result while the dashed line shows
the approximate solution (\ref{sol_son}), rescaled by an overall
factor $c$, $\Delta(p_0)=c\Delta_{\rm app.}(p_0/c)$. At $\mu=10^{10}$ 
MeV, $g\simeq 0.67$ and Son's solution is in excellent agreement
with the exact result, up to an overall factor $c\simeq 2$. At
$\mu=400$ MeV the coupling is significantly bigger than 1, $g
\simeq 3.43$, but the approximate solution is still qualitatively 
correct. 

 The scaling of the maximum gap with the chemical potential is shown
in Fig. \ref{fig_scale}. The solid line is the numerical result 
and the dashed lines correspond to $cg^{-k}\mu\exp(-3\pi^2/(\sqrt{2}g))$
with $k=0,\ldots,5$. We observe that the $k=0$ curve provides an 
excellent fit to the data even for small chemical potentials. 
Again, the overall coefficient is $c\simeq 2$. 

 Let us make a few observations at this point. First, we note that
the use of perturbation theory to determine the dynamic screening
is self consistent. Since $\Delta\sim\mu\exp(-{\rm const.}/g)$, the gap
grows as $\mu\to\infty$ and $q_M\gg\Lambda_{QCD}$. Second, we note
that it is essential to keep the frequency dependence of the gap. 
For small frequencies $\Delta(p_0)$ varies over scales on the 
order of $p_0\sim\Delta_0$ itself. Therefore, $\Delta(p_0)$ cannot
be replaced by a constant.  Were we to approximate $\Delta(p_0)
\simeq\Delta_0$, as in \cite{Hon_99},
we would obtain a gap equation for $\Delta_0$ that has
the correct double logarithmic structure and gives $\Delta_0
\simeq\mu\exp(-{\rm const.}/g)$, but the constant in the exponent
would  not be correct. 

 Finally, we note that it is easy to see what taking into account
the numerical coefficients and the factor $g^2$ in equation 
(\ref{eliash}) will do. Any numerical factor inside the logarithm
can be absorbed by rescaling the frequencies. Therefore, if 
$\Delta_{\rm app.}(p_0)$ in (\ref{sol_son}) is an approximate solution 
to (\ref{eliash1}), then $\Delta'(p_0)=c\Delta_{\rm app.}(p_0/c)$ with
$c=64\pi/(N_fg^2)$ is an approximate solution to (\ref{eliash}).
This can also be seen from Figs. \ref{fig_gap_mag} and 
\ref{fig_scale_mag} where we show the numerical solution to the
Eliashberg equation (\ref{eliash}) for the superconducting 
gap from magnetic gluon exchanges. Asymptotically, the solution
is well described by the function $\Delta'(p_0)$ with $c\simeq
175 g^{-2}$. 

 We now come to the role of electric gluon exchanges. We include
the second term in (\ref{gap_5}) with $F=m_D^2$. We again use the
approximation $\cos\theta\simeq 1$ in the numerator and drop 
the $(q_0-p_0)^2$ term in the denominator. Let us note that 
in the forward direction, electric and magnetic gluon exchanges
have the same overall factor. Performing the integral over
$\cos\theta$, we find
\be
\label{eliash_mel}
\Delta(p_0) = \frac{g^2}{18\pi^2} \int dq_0
 \left\{ \log\left(1+\frac{64\pi\mu}{N_fg^2|p_0-q_0|}\right)
     +  \frac{3}{2}\log\left(1+\frac{8\pi^2}{N_fg^2} \right) \right\} 
 \frac{\Delta(q_0)}{\sqrt{q_0^2+\Delta(q_0)^2}},
\ee
where the factor $3/2$ in front of the second term comes from the
difference between dynamic screening, $q_M\sim |\vec{q}|^{1/3}$, and
static screening, $q_E\sim |\vec{q}|$. In the weak coupling limit
we again expect the solution to be of the form $c\Delta_{\rm app}(p_0/c)$
with 
\be 
 c = 1024\sqrt{2}\pi^4 N_f^{-5/2}\, g^{-5} 
   = 256\pi^4 g^{-5}\simeq 2.5\cdot 10^4 g^{-5}
   \hspace{0.5cm} (N_f=2).
\ee
We can compare this prediction to our numerical results, obtained
from solving (\ref{gap_5}). In this equation, we take into account
both electric and magnetic gluon exchanges. We also keep the 
$\cos\theta$ dependence in the numerator, and the terms $(q_0-p_0)^2$
in the denominator. Finally, we use the exact form of $G$ and $F$
in the hard dense loop approximation, 
\bea
F&=&2m^2\frac{q^2}{\vec{q}^{\,2}}\Bigg( 1-\frac{iq_0}{|\vec{q}|}
     Q_0\Bigg(\frac{iq_0}{|\vec{q}|}\Bigg)\Bigg),
\hspace{1cm} Q_0(x)=\frac{1}{2}\log\left(\frac{x+1}{x-1}\right), \\
G&=&m^2\frac{iq_0}{|\vec{q}|}\Bigg[
    \Bigg(1-\Bigg(\frac{iq_0}{|\vec{q}|}\Bigg)^2\Bigg)
   Q_0\Bigg(\frac{iq_0}{|\vec{q}|}\Bigg) +
   \frac{iq_0}{|\vec{q}|} \Bigg].
\eea
This takes into account that there is no dynamic screening
for $|\vec{q}|<q_0$. 
The numerical results are shown in Figs. \ref{fig_gap_full} and
\ref{fig_scale_full}. Asymptotically, the gap is well described 
by $c\Delta_{\rm app}(p_0/c)$ with $c\simeq 1.4\cdot 10^4 g^{-5}$. 
We notice that for $\mu=10^{10}$ MeV the solution has a `knee'
at $p_0\simeq 10^9$ MeV. This comes from the fact that for 
frequencies $p_0>\sqrt{N_f/(8\pi)}g\mu$ the retardation terms
$\sim(p_0-q_0)^2$ dominate over screening. In this regime, the
solution is of the same form, but the scale factor is different.

 Overall, the scaling with $g^{-5}\exp(-3\pi^2/(\sqrt{2}g))$
is clearly visible, though not quite as impressively as in the case
with magnetic gluon exchange only. For chemical potentials
that are of physical interest, $\mu<1000$ MeV, the gap reaches
$\Delta_0\simeq 100$ MeV. We should caution, however, that 
in this regime $g\simeq (2-4)$, and higher order corrections
are probably important. Nevertheless, it is gratifying to see
that the order of magnitude of the result agrees with 
previous calculations \cite{ARW_98,RSSV_98} 
based on more phenomenological 
effective interactions, which were normalized 
to the strength of chiral symmetry
breaking at zero density, rather than the calculable asymptotics of
the running coupling. 

 4. There are a number of questions that will need to be addressed 
in a more complete calculation. First, we have concentrated 
on the case $N_f=2$. For $N_f=3$, there are two order parameters,
corresponding to the color antisymmetric and color symmetric 
components of the color-flavor locked state. This is only
a minor complication, since there is only one combination that
survives in the weak coupling limit.

 A more complicated issue is the role of the Meissner effect.
For $N_f=2$, the dominant order parameter only breaks color
$SU(3)\to SU(2)$, and all gluons that contribute to pairing,
except for one, live in the unbroken part of the gauge group. In 
the case of $N_f=3$, the Higgs mechanism is complete and all gluons
acquire a mass. At zero momentum and frequency, the screening 
mass is on the order of $m^2\sim g^2\mu^2$, much larger than the 
dynamic screening scale $q_M$. At finite momentum transfer, on the 
other hand, the screening mass is $m^2\sim g^2\mu^2\Delta/|\vec{q}|$
\cite{AGD}, which is of the same form as the dynamic screening 
effect. The Meissner effect will therefore not affect the 
dependence of the gap on the coupling constant, but it will
affect the numerical coefficient. 

 Finally, one has to address higher order corrections to 
the perturbative result. In particular, one would like to 
know what the functional form of the corrections is, and
whether the applicability of perturbation theory requires 
$g<1$, or some weaker condition like $g<\pi$. We have 
already mentioned wave function renormalization
as one source of higher order correction \cite{Son_98}. 
Another issue is vertex corrections. The vertex correction
generated by hard dense loops is \cite{LeB}
\be
\label{vertex}
 \Gamma_\mu^a(p_1,p_2) = g\frac{\lambda^a}{2}
 \left( \gamma_\mu -m_f^2 \int \frac{d\Omega}{4\pi}
  \frac{\hat{K}_\mu \gamma\cdot\hat{K}}{(p_1\cdot\hat{K})
   (p_2\cdot\hat{K})} \right),
\ee
where $\hat{K}=(i,\hat{k})$ is a light like vector and
$m_f^2=g^2\mu^2/(6\pi^2)$. We can insert this correction 
into the gap equation (\ref{gap3}). We find that the 
coefficient of the magnetic gluon exchange is modified
as
\be
\label{v_cor}
 \left(\frac{3}{2}-\frac{1}{2}\cos\theta\right)
\to \left(\frac{3}{2}-\frac{1}{2}\cos\theta\right)
 + \frac{m_f^2}{\mu^2} +\frac{1}{2}(1-\cos\theta)
 m_f^2\int\frac{d\Omega}{4\pi}\frac{1}{(p\cdot\hat{K})
 (q\cdot\hat{K})}.
\ee
In the forward direction $\cos\theta\simeq 1$, which 
dominates the gap equations, this is just a higher order 
correction. Vertex corrections have a $(\mu/q_0)$ 
enhancement in the backward direction $\cos\theta\simeq
-1$, but the integral over $\cos\theta$ is finite as
$q_0\to 0$. Vertex corrections in the magnetic part 
therefore do not modify the asymptotic form of the gap.
The same is true for vertex corrections in the electric 
part of the interaction. 

 5. In summary, we have performed a perturbative 
calculation of the superconducting gap in two flavor
QCD at very high density. We find that the gap scales
as $\Delta_0\simeq 256\pi^4\mu g^{-5}\exp(-3\pi^2/
(\sqrt{2}g))$, where the overall coefficient is correct 
up to a factor of order one. In the physically interesting 
regime $\mu<1$ GeV, the gap is on the order of 100 MeV, 
in agreement with earlier calculations based on 
instantons or schematic interactions adjusted to
the size of the chiral condensate at zero density.

 Acknowledgments: We would like to thank D. Son and
K. Rajagopal for useful discussions.

% figures

\newpage\noindent

\begin{figure}
\caption{\label{fig_gap}
Solution of the Eliashberg equation (\ref{eliash1}) as a 
function of imaginary frequency $q_0$. The upper and lower 
panels show the solution for $\mu=400$ MeV and $\mu=10^{10}$
MeV, respectively. The solid lines show the numerical solution 
and the dashed lines shows the approximate solution (\ref{sol_son}), 
scaled to the same value of the gap.}
\end{figure}

\begin{figure}
\caption{\label{fig_scale}
Dependence of the gap on the chemical potential for
the solution of the Eliashberg equation (\ref{eliash1}).
Here, $g(\mu)$ is taken to run according to the one-loop
beta function. The dotted curves show the functions 
$g^{-k}\exp(-(3\pi^2)/(\sqrt{2}g))$ for $k=0$ (top),
$\ldots,5$ scaled to the value of the gap at the maximum 
chemical potential.}
\end{figure}

\begin{figure}
\caption{\label{fig_gap_mag}
Same as figure \ref{fig_gap} for the solution of the 
Eliashberg equation with magnetic gluon exchange only,
see equ. (\ref{eliash}).}
\end{figure}

\begin{figure}
\caption{\label{fig_scale_mag}
Same as figure \ref{fig_scale} for the solution of the 
Eliashberg equation with magnetic gluon exchange only.}
\end{figure}

\begin{figure}
\caption{\label{fig_gap_full}
Same as figure \ref{fig_gap} for the solution of the 
Eliashberg equation with magnetic and electric gluon 
exchanges, see equ. (\ref{gap_5}).}
\end{figure}

\begin{figure}
\caption{\label{fig_scale_full}
Same as figure \ref{fig_scale} for the solution of the 
Eliashberg equation with magnetic and electric gluon exchanges.}
\end{figure}

\newpage\noindent
\setcounter{figure}{0}

\begin{figure}
\begin{center}
\leavevmode
\vspace{-1cm}
\epsfxsize=10cm
\epsffile{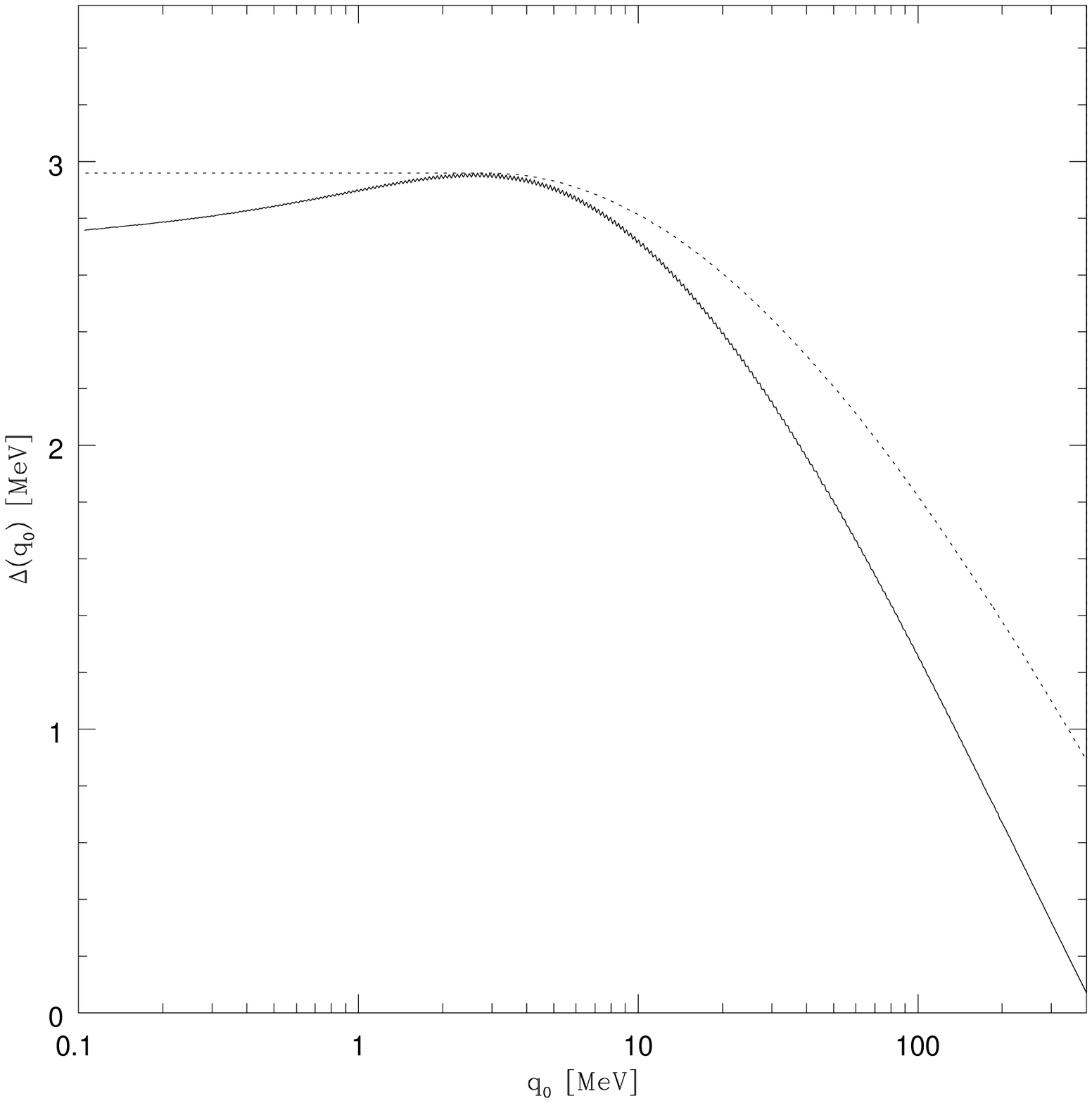}
\end{center}\begin{center}
\leavevmode
\epsfxsize=10cm
\epsffile{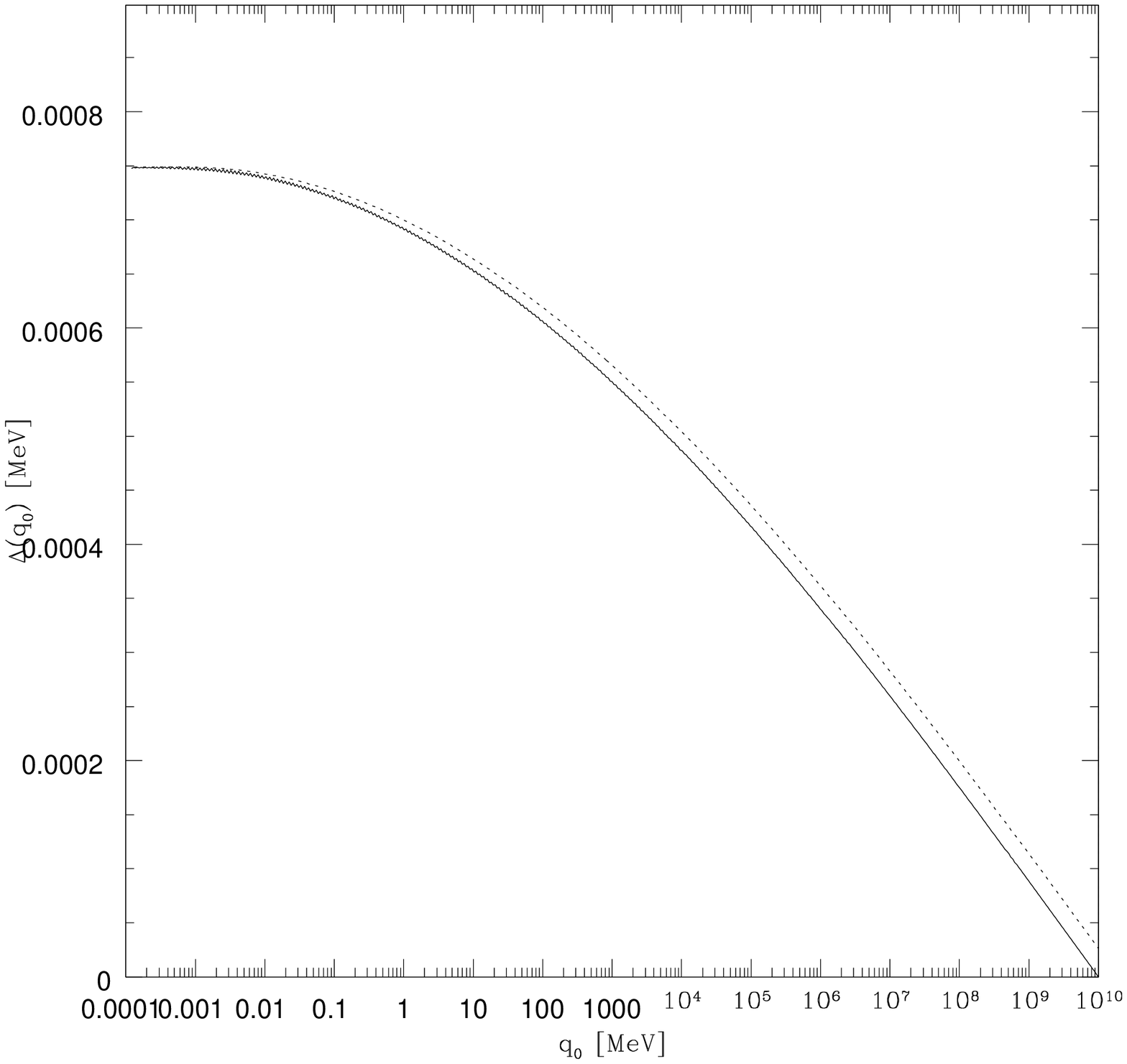}
\end{center}
\vspace*{-1cm}\caption{}
\end{figure}

\begin{figure}
\begin{center}
\leavevmode
\vspace{0.5cm}
\epsfxsize=14cm
\epsffile{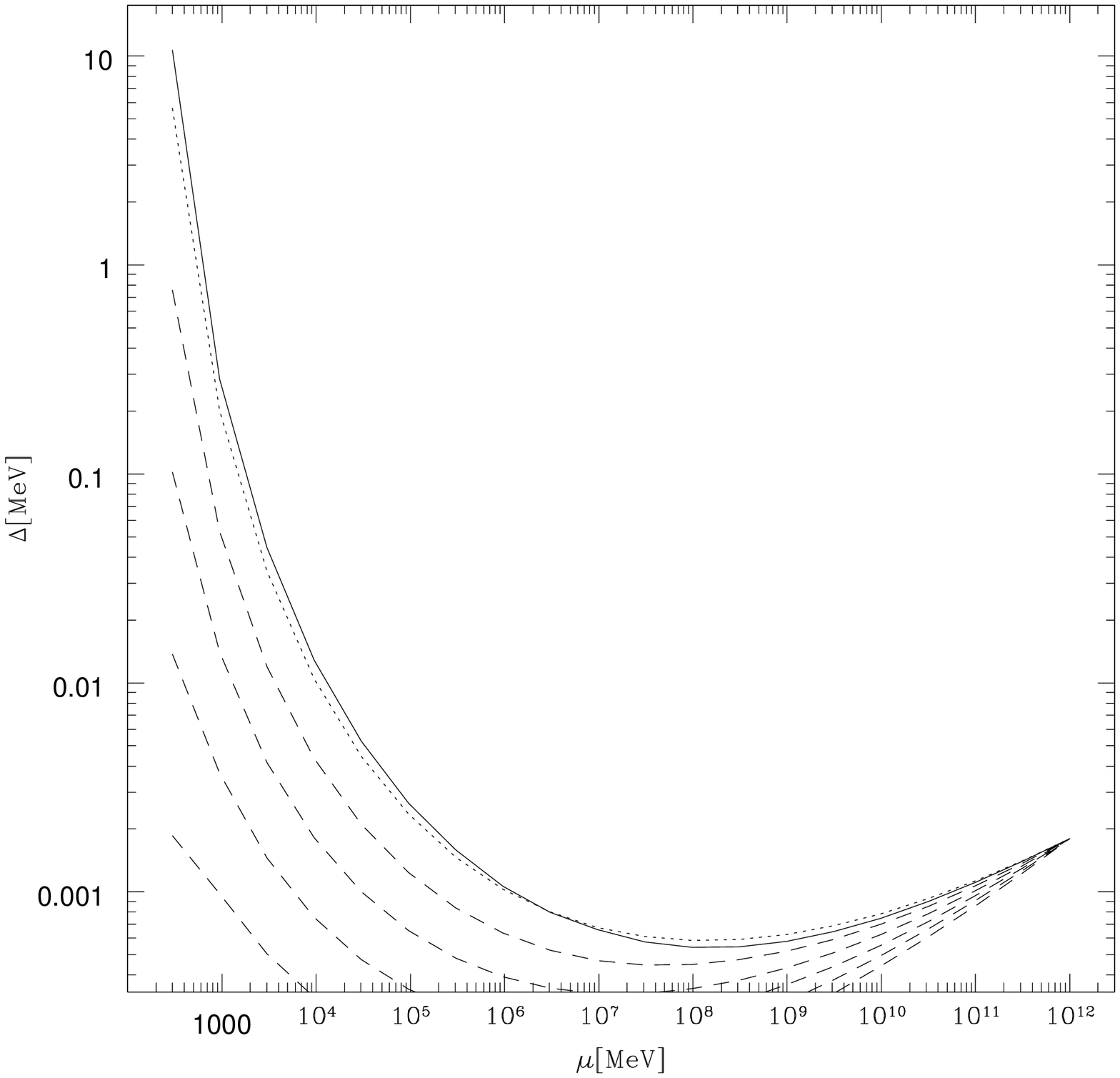}
\end{center}
\vspace*{-1cm}

\caption{}
\end{figure}

\begin{figure}
\begin{center}
\leavevmode
\vspace{-1cm}
\epsfxsize=10cm
\epsffile{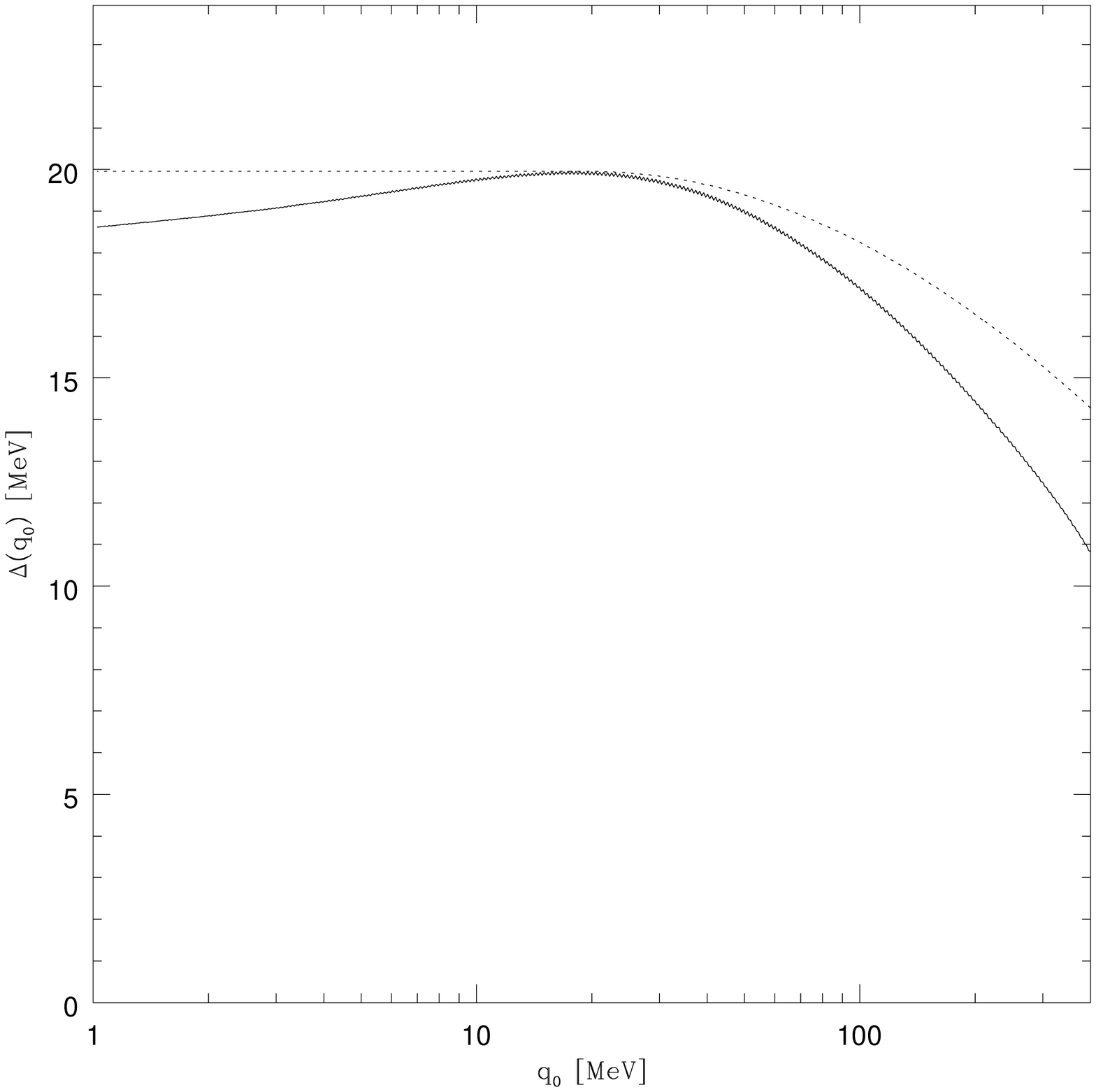}
\end{center}\begin{center}
\leavevmode
\epsfxsize=10cm
\epsffile{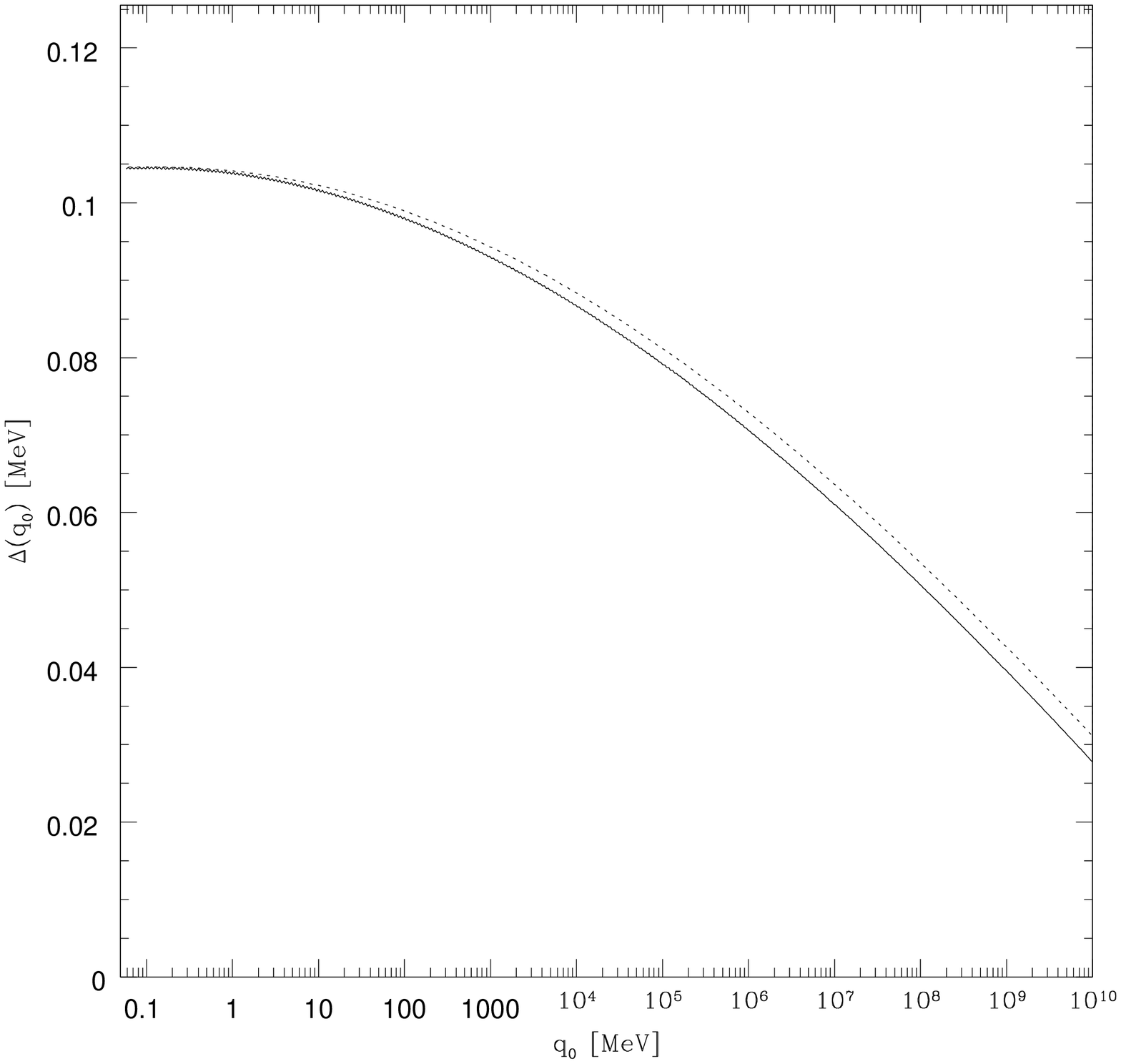}
\end{center}
\vspace*{-1cm}
\caption{}
\end{figure}

\begin{figure}
\begin{center}
\leavevmode
\vspace{0.5cm}
\epsfxsize=14cm
\epsffile{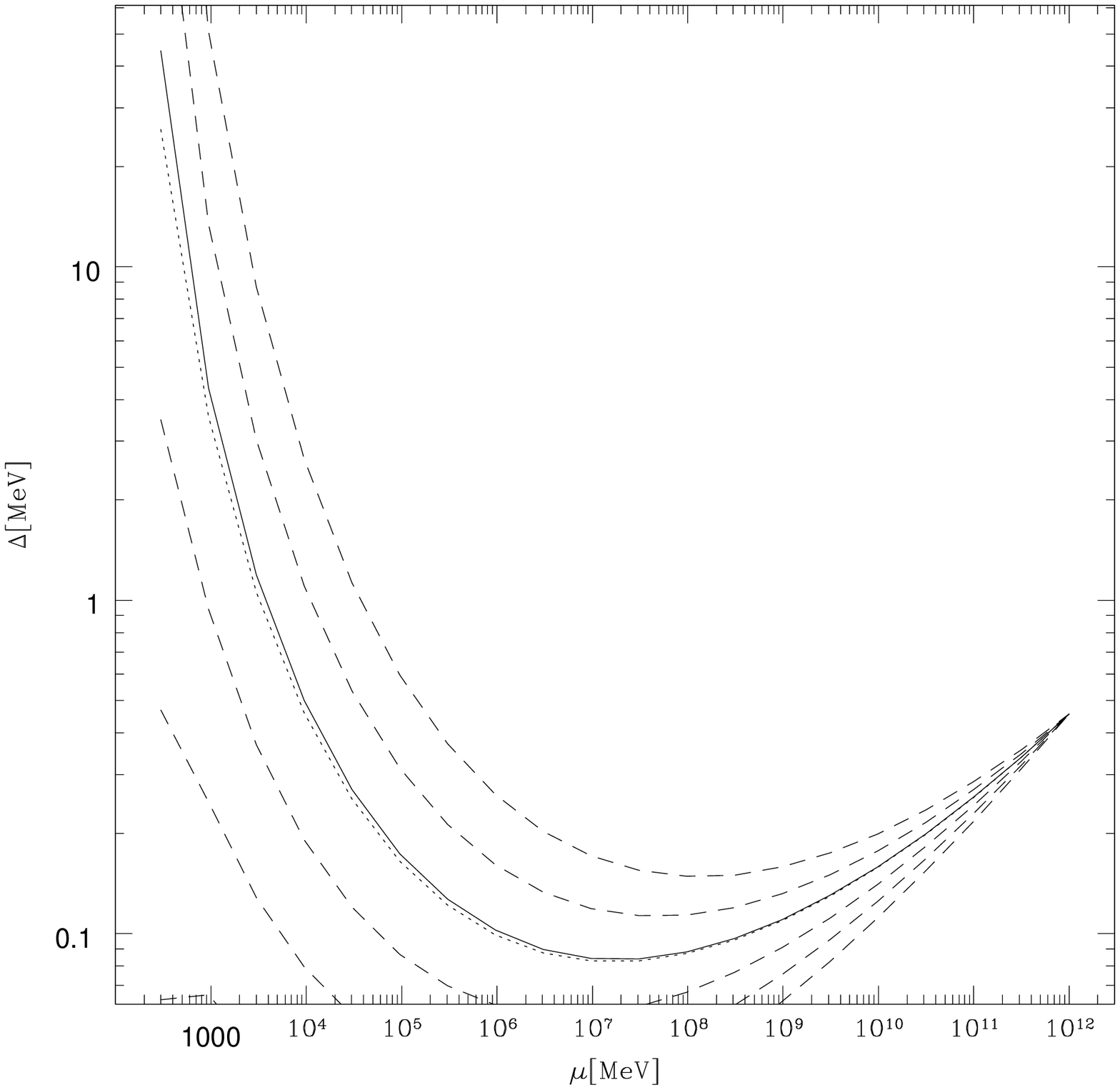}
\end{center}
\vspace*{-1cm}
\caption{}
\end{figure}

\begin{figure}
\begin{center}
\leavevmode
\vspace{-1cm}
\epsfxsize=10cm
\epsffile{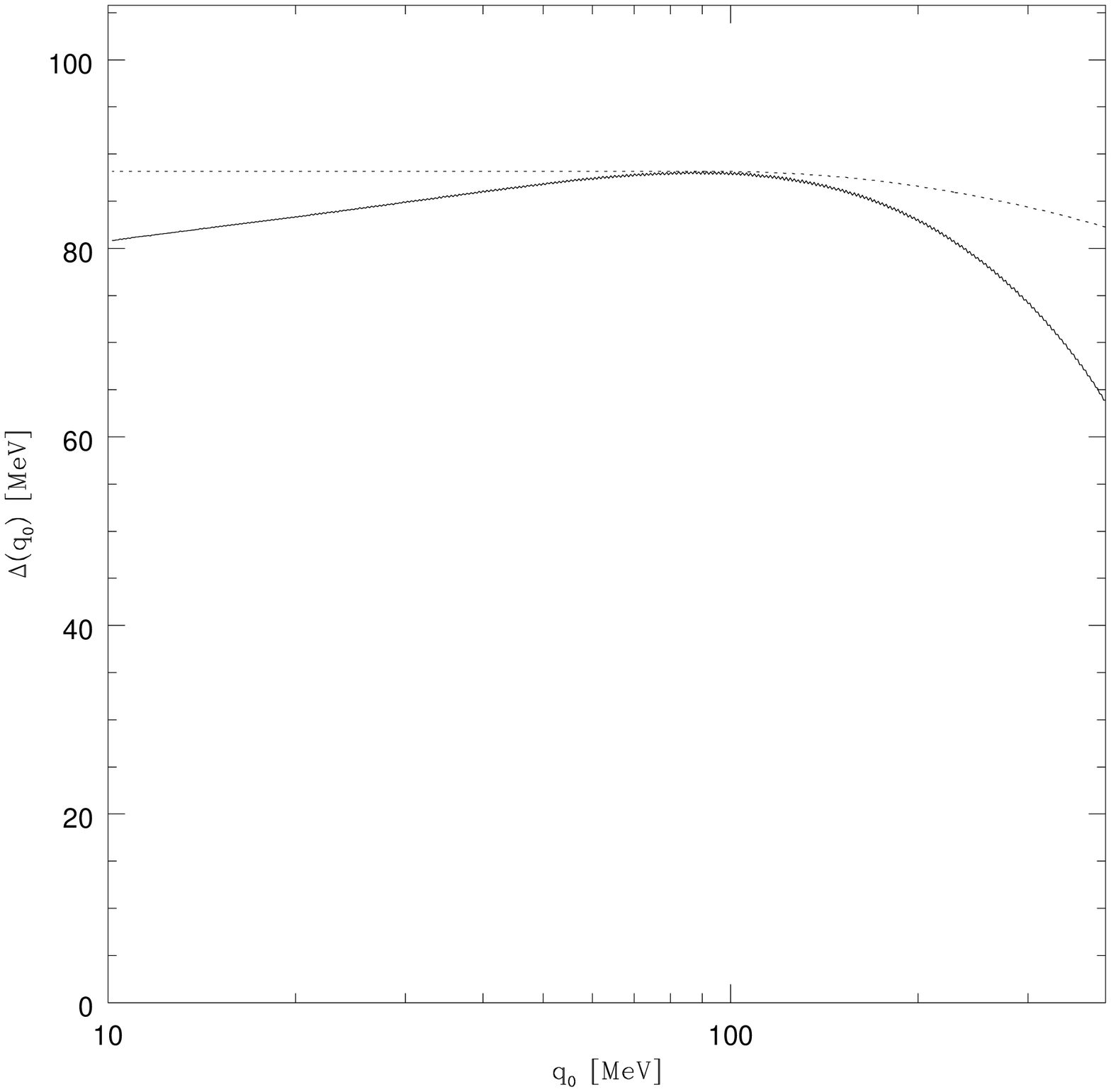}
\end{center}\begin{center}
\leavevmode
\epsfxsize=10cm
\epsffile{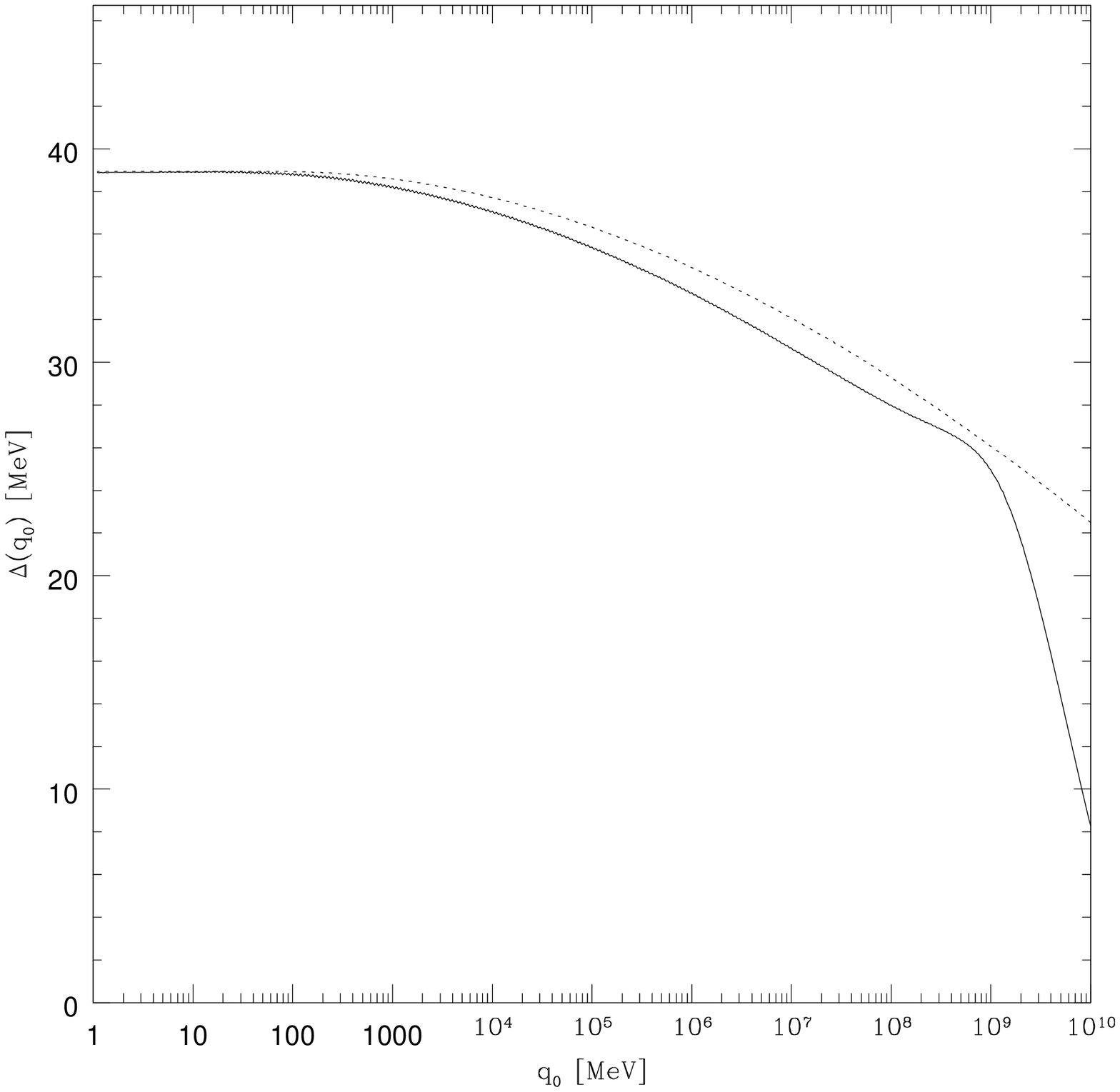}
\end{center}
\vspace*{-1cm}
\caption{}
\end{figure}

\begin{figure}
\begin{center}
%\leavevmode
\vspace{0.5cm}
\epsfxsize=14cm
\epsffile{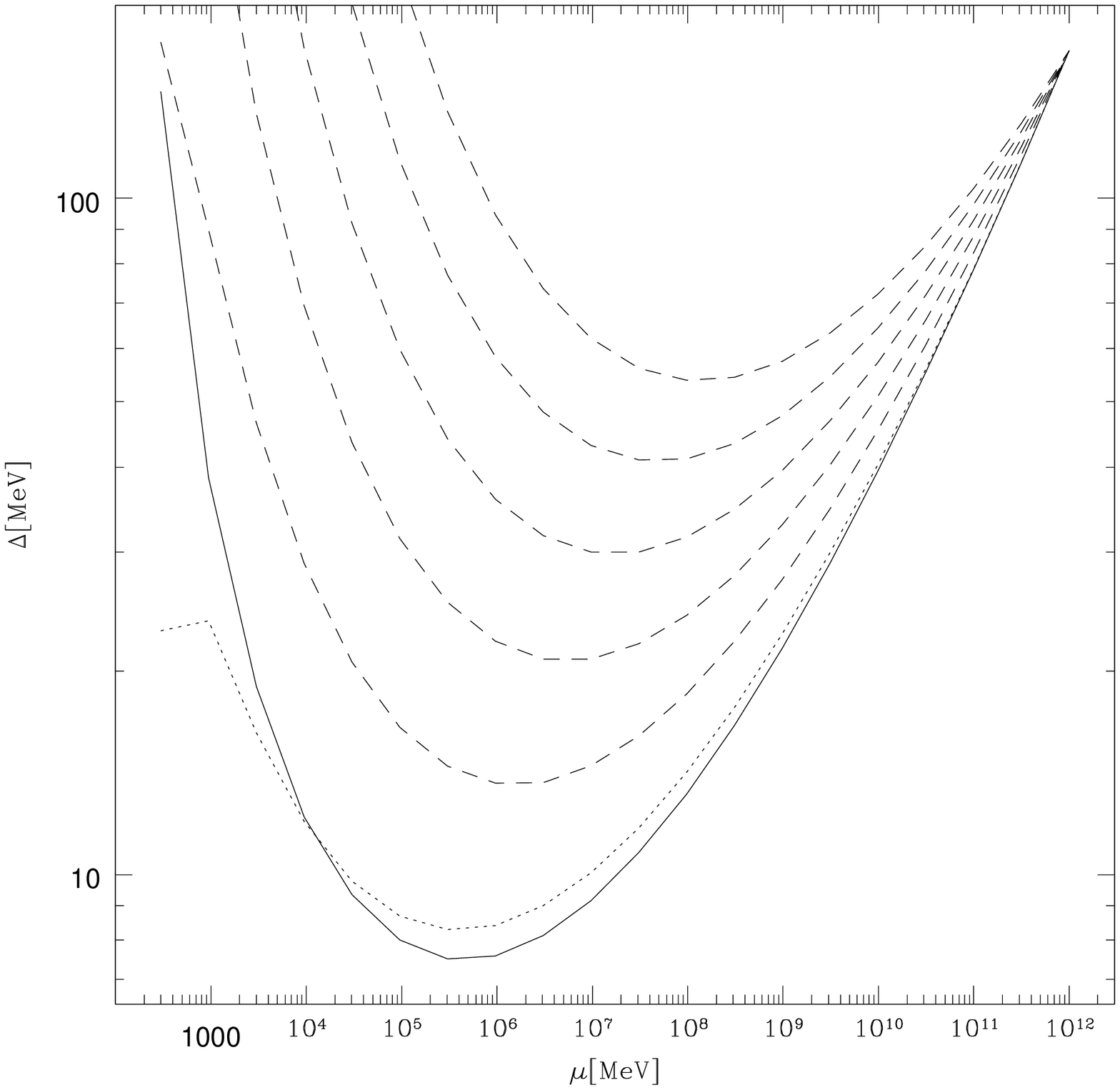}
\end{center}
\vspace*{-1cm}
\caption{}
\end{figure}

\end{document}